\documentclass[conference]{IEEEtran}
\IEEEoverridecommandlockouts

\usepackage{cite}
\usepackage{amsmath,amssymb,amsfonts}
\usepackage{algorithmic}
\usepackage{mhchem}
\usepackage{graphicx}
\usepackage{textcomp}
\usepackage{xcolor}
\usepackage{booktabs}
\usepackage[hidelinks]{hyperref}
\usepackage{siunitx}
\def\BibTeX{{\rm B\kern-.05em{\sc i\kern-.025em b}\kern-.08em
    T\kern-.1667em\lower.7ex\hbox{E}\kern-.125emX}}
\begin{document}
\title{Demand response potential evaluation of a zero carbon hydrogen metallurgy system considering shaft furnace's flexibility
\thanks{This work was supported  By Science and Technology Program of Xinjiang Uyghur Autonomous Region (No. 2024B01001), *corresponding author: Yue Zhou). 
\vspace{-1.5em}
}
}

\author{\IEEEauthorblockN{Qiang Ji, Lin Cheng, Kaidi Huang, Junxin Lv}
\IEEEauthorblockA{\textit{dept. of Electrical Engineering} \\
\textit{Tsinghua University}\\
Beijing, China \\
jiq23@mails.tsinghua.edu.cn
}
\and
\IEEEauthorblockN{Yue Zhou*}
\IEEEauthorblockA{\textit{State Key Laboratory of Smart}\\ \textit{Power Distribution Equipment and System}\\
\textit{Tianjin University}\\
Tianjin, China \\
zhouyue68@tju.edu.cn
}
\and
\IEEEauthorblockN{Zeng Liang}
\IEEEauthorblockA{\textit{School of Metallurgical and} \\ \textit{Ecological Engineering} \\
\textit{University of Science}\\
and Technology Beijing \\}
\textit{Beijing, China} \\
d202210160@xs.ustb.edu.cn
}

\maketitle
\begin{abstract}
The increasing penetration of intermittent renewable energy sources and the retirement of thermal units have widened the power system flexibility gap. Industrial demand response (DR) driven by real-time pricing is widely regarded as a viable solution. In this paper, we propose a framework to quantify the DR potential of a zero-carbon hydrogen metallurgy system (ZCHMS) considering shaft furnace's flexibility. First, we model the shaft furnace as a constrained flexible load and validate the model via simulation, achieving a root mean square error of 4.48\% of the rated load. Second, we formulate a DR potential evaluation method that determines baseline and DR-based production scheduling schemes by minimizing operating cost subject to production orders. Finally, the numerical results show that compared with the baseline, DR-based ZCHMS reduces operating cost by 6.6\%, incentivizing demand-side management in ironmaking and strengthening power-ironmaking synergies.
\end{abstract}
\begin{IEEEkeywords}
Demand response, direct reduced iron, flexibility, shaft furnace, zero-carbon, heavy industry.
\end{IEEEkeywords}
\section{Introduction}\label{Introduction}
The increasing penetration of intermittent renewable energy sources (RES) and the retirement of thermal units have widened the power system flexibility gap \cite{yu2023demand}. Demand response (DR) induces end-users to reshape their consumption patterns in response to real-time pricing and incentive subsidies, thereby smoothing peak-valley differences and providing flexibility reserves for the power system \cite{li2023coordinated}. Among all consumer sectors, industrial facilities represent the largest DR potential \cite{golmohamadi2022demand}, particularly in a zero carbon hydrogen metallurgy system (ZCHMS) for ironmaking, which enables deep decarbonization of ironmaking and absorbs surplus RES \cite{ji2025energy}. Therefore, quantifying and harnessing ZCHMS's flexibility through DR is crucial for reducing electricity costs and strengthening power-ironmaking synergies \cite{lei2023,li2025redesigning}. \par 
Most studies on hydrogen-based shaft furnace (SF) ironmaking focus on improving energy efficiency and reducing \ce{CO2} emissions by adopting top-gas recycling \cite{zhang16}, low-temperature waste heat recovery \cite{ji2025energy}, and electric heating to displace combustion. Because of product-quality and safety constraints, these studies model the alkaline electrolyzer (AE) and the SF as steady-state units (i.e., the AE operates at fixed power and the SF maintains at a constant discharge flow rate). Consequently, the ZCHMS lacks the ability to effectively respond to RES volatility, leading to significant wind and photovoltaic (PV) curtailment as well as increased operating costs. In contrast, some studies model the SF as high-level "flexible" load to minimize the operation costs of the ZCHMS \cite{wu2025,sheng2024}, neglecting the dynamic quasi-steady-state conditions of SF and the influence of transients on the metallization rate of direct reduced iron (DRI). This oversimplification may lead to DRI with a metallization rate outside the acceptable range, compromising product quality. Therefore, it is essential to explicitly model the SF's flexibility and further analyze the DRI metallization rate under dynamic operation, thereby enhancing the flexibility resources available within ZCHMS. \par
Moreover, most industrial DR papers concentrate on oil refineries \cite{alarfaj2018material}, cement manufacturing and aluminum smelting plants \cite{golmohamadi2022demand}, chemical plants \cite{otashu2019demand}, electric arc furnace or ladle furnace \cite{wang2023quantifying}. However, little attention has been paid to evaluating the DR potential of ZCHMS, limiting the utilization of its flexibility to coping with RES and electricity prices variability. \par
To bridge the research gaps, we propose a framework to quantify the DR potential of the ZCHMS considering SF's flexibility. The main contributions of this paper are summarized as follows: \par
1) The dynamic regulation process of SF is modeled as a 1-order transition process at an hourly resolution. The parameter is identified using physics-based dynamic operating data, achieving a root mean square error (RMSE) of 4.48\% of the rated load. Additionally, the metallization rate of DRI is changed within safety limits, satisfying product-quality specifications.\par
2) We formulate a DR potential evaluation method that determines baseline and DR-based production scheduling schemes. Compared with the baseline, DR-based the ZCHMS reduces operating costs by 6.6\%, demonstrating the effectiveness and advancement of the proposed framework.
\begin{figure*}[t!]
\centering
\includegraphics[width=1\textwidth]{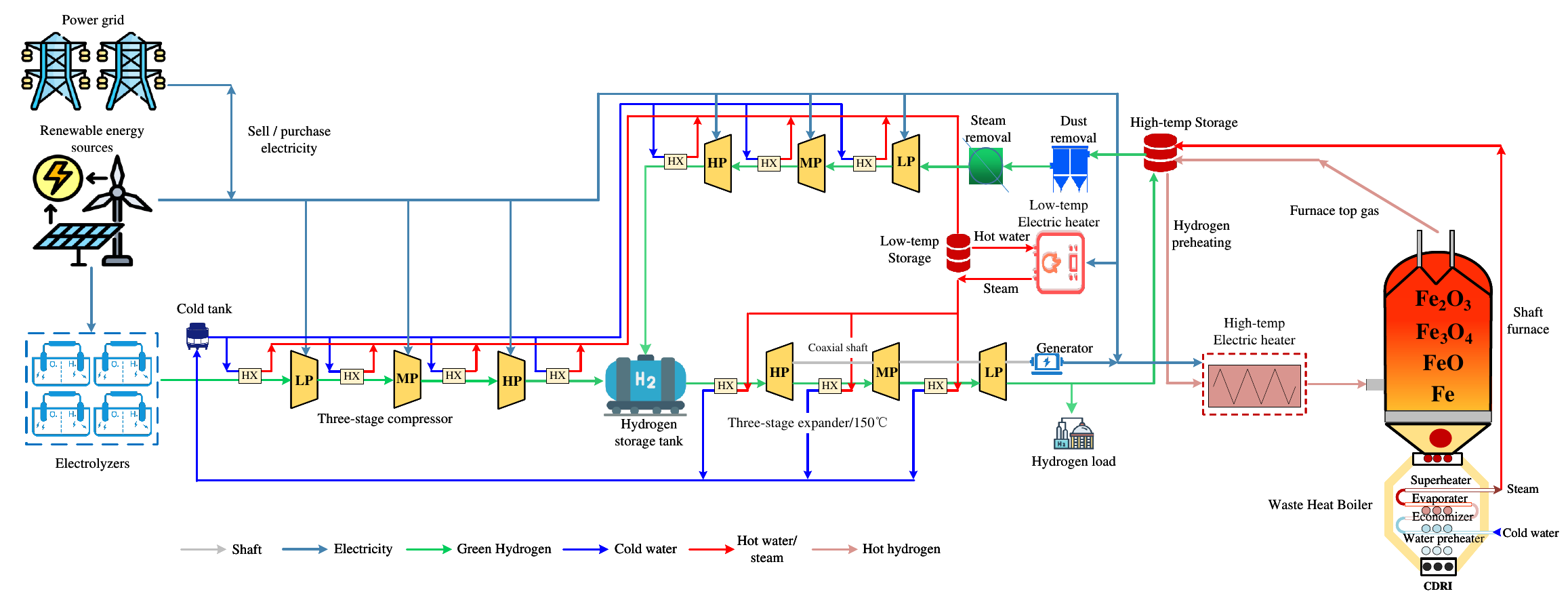}
\setlength{\abovecaptionskip}{-0.1cm}  
\setlength{\belowcaptionskip}{-0.1cm} 
\caption{The zero-carbon hydrogen metallurgy production system}
\vspace{-0.1 cm}
\end{figure*}
\par The rest paper is organized as follows: Section 2 describes the production process of ZCHMS and limited flexibility model of SF. The DR potential evaluation framework is introduced in Section 3. The case study is carried out in Section 4. Conclusions are drawn in Section 5.
\section{The production process of ZCHMS and limited flexibility model of SF}
\subsection{The production process of ZCHMS}
\par The production process of ZCHMS is shown in {\color{blue}Fig. 1}. The power side includes wind and PV generation, a three-stage expander, and the external grid. The load side includes compressors, AE, high- and low-temperature electric resistance heaters, and the SF. The storage subsystem includes high- and low-temperature thermal storage tanks and hydrogen storage tank. Given the production orders and the SF's complex start-up/shut-down characteristics, this paper does not consider interruptible loads; instead, sheddable and shiftable loads are evaluated  during the DR period. 
\subsection{Limited flexibility model of SF}
When the internal temperature of the SF exceeds 570°C, the reduction hydrogen undergoes a three-stage process as it rises from the medium to the top of the SF, ultimately yielding DRI. The detailed energy-mass balance can be found in \cite{ji2025energy}. The consumption rate of hydrogen $Md_{{H_2},out}^t$ is as follows:
\begin{equation}
   Md_{{H_2},out}^t = {\phi _{H_2}}M_{DRI}^t
\end{equation}
$\phi_{H_2}$ is the \ce{H2}-DRI coefficient; $M_{DRI}^t$ is the discharge flow rate of SF. \par 
The dynamic regulation process of SF must account for internal reactions and pressurization of discharge bin. Therefore, we adopt a one-hour time step for the flexibility model. Within each scheduling period interval, the internal physico-chemical processes of SF is considered to reach a quasi-steady state. The limited flexibility model of SF is proposed in eq. (2). We employ the common 1-order transition process in chemical process as a reference to characterize the discrete model by eq. (3), the variable load range of SF in eq.~\eqref{Ra1}, and ramping limitations in eq.~\eqref{Rp1}, denoted as follows: 
\begin{equation}
\label{eq_M1}
\begin{split}
    M_{DRI}^{t + 1} = \alpha M_{DRI,QSS}^t + (1 - \alpha)M_{DRI,QSS}^{t + 1}, \alpha=e^{-\frac{t}{T_{trans}}}
\end{split}
\end{equation}
\label{K_1}
\begin{equation}
    M_{DRI}^{k + 1} = M_{DRI,QSS}^{(k + 1)} + (M_{DRI,QSS}^k - M_{DRI,QSS}^{(k + 1)}){e^{ - \frac{t}{{{T_{trans}}}}}}
\end{equation}
\begin{equation}
\label{Ra1}
    M_{DRI,dis}^{\min } \le M_{DRI,QSS}^k \le M_{DRI,dis}^{\max }
\end{equation}
\begin{equation}
\label{Rp1}
    \Delta M_{DRI,dis}^-  \le M_{DRI,QSS}^{k + 1} - M_{DRI,QSS}^k \le \Delta M_{DRI,dis}^+ 
\end{equation}
$\alpha$ is the lag factor; $T_{trans}$ is the transition time constant; $M_{DRI,QSS}^{k+1}$ is $k$\textit{th} quasi-steady-state condition of the SF; $M_{DRI,dis}^{\min},M_{DRI,dis}^{\max}$ are the lower and upper bounds of discharge rate of DRI, respectively; $M_{DRI,dis}^-,M_{DRI,dis}^+$ are the lower and upper bounds of ramping rate, respectively. \par
Additionally, we adopt a validated dynamic model \cite{immonen2024dynamic} to simulate actual regulation processes of SF. The control strategy first sets the hydrogen feed rate according to the requirement of the next-step DRI reduction and then adjusts the discharge flow rate of DRI. The adjustment is considered settled once the metallization rate of DRI remains within ±0.5\% of the setpoint. The simulation results of SF are presented in Fig. 2, where the discharge rate of the SF increases from 30t/h to its rated value of 125t/h. Subsequently, the parameter of the proposed flexibility model is identified, and yielding a transition time constant of $T_{trans} = 1.27$ h. The dynamic simulation and the identified model simulation are both plotted in Fig. 2(a), with a root mean square error of 6.73t/h (4.48\% of rated load). Moreover, the metallization rate of DRI is changed within safety limits, satisfying product-quality specifications, shown in Fig. 2(b). Thus, the proposed flexibility model of SF exhibits high accuracy and suitability for optimization-based scheduling. 
\begin{figure}[htbp] 
    \setlength{\abovecaptionskip}{-0.1cm}  
    \setlength{\belowcaptionskip}{-0.1cm}   
    \includegraphics[width=1\columnwidth]{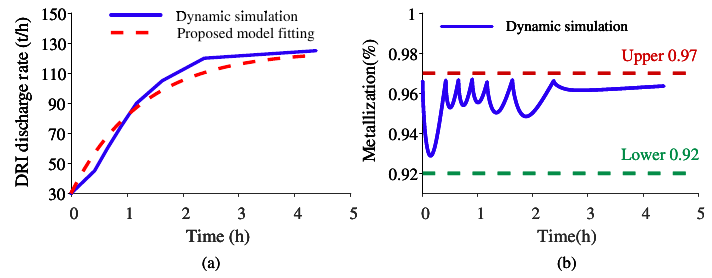}
    \caption{Dynamic regulation processes of SF under simulation data and the identifier model. (a) Discharge flow rate of DRI. (b) Metallization rate of DRI}
    \label{core diagram}
\end{figure}
\subsection{The components operation models of ZCHMS}
1) Alkaline electrolyzer: AE $P_{AE}^{t}$ is used to produce hydrogen $M_{\ce{H2},in}^{t}$ via water electrolysis. The low-temperature waste heat recovery model $W_{AE}^{t}$ and the constant efficiency model are shown in eq.~\eqref{W_AE} and in eq.~\eqref{P_AE}, respectively.The variation load-range constraint is adopted in eq.~\eqref{P_min_max}.
\begin{equation}
\label{W_AE}
    W_{AE}^t = {\gamma _{{H_2}}}M_{{H_2},in}^t
\end{equation}
\begin{equation}
\label{P_AE}
    P_{AE}^t = {\kappa _{{H_2}}}M_{{H_2},in}^t
\end{equation}
\begin{equation}
\label{P_min_max}
    C_{AE}^{\min } \le P_{AE}^t \le C_{AE}^{\max}
\end{equation}
\begin{equation}
    {C_{AE}} = {N_{AE}}C_{AE}^{single}
\end{equation}
$\gamma _{{H_2}}, \kappa _{{H_2}}$ are low-temperature waste heat coefficient and the energy efficiency coefficient of the AE, respectively. $C_{AE}^{\min },C_{AE}^{\max}$ are the lower and upper bounds of single AE, respectively; $N_{AE}$ is the total number of AE. \par
2) Hydrogen storage tank: the hydrogen storage tank is used to smooth fluctuations in RES output and electricity prices. During charging, a three-stage compressor $P_{comp}^{t}$ is employed and the low-temperature waste heat $W_{comp}^{t}$ generated in compression is recovered. During discharging, a three-stage expander $P_{exp}^{t}$ is used to generate electricity, with hydrogen preheating $W_{exp}^{t}$ incorporated into the process.
\begin{equation}
    H{T^{t + 1}} = H{T^t} + (M_{{H_2},in}^t - M_{{H_2},out}^t)
\end{equation}
\begin{equation}
    M_{{H_2},out}^t = Md_{{H_2},out}^t + Ml_{sell}^t
\end{equation}
\begin{equation}
    Ml_{sell}^t = {H_2}^{request}
\end{equation}
\begin{equation}
    H{T_{\min }} \le H{T_t} \le H{T_{\max }}
\end{equation}
\begin{equation}
    HT^1 = HT^{T+1} = 50\% HT
\end{equation}
$Ml_{sell}^{t}$ is the hydrogen load; $M_{H_{2},out}^{t}$ is the discharge hydrogen rate; $HT_{min}, HT_{max}$ are the lower and upper bounds of hydrogen storage tank.\par
3) High-temperature electric heater (Hteh): Hteh $P_{Eh}^{t}$ is used to heat the hydrogen to 1050℃, supplying the reaction heat flow $W_{sf,H_2}^{in}$.
\begin{equation}
    W_{Eh}^{in} = P_{Eh}^t\Delta t
\end{equation}
\begin{equation}
    W_{Eh}^{out} = W_{Eh}^{in}{\phi _{Eh}}
\end{equation}
$W_{Eh}^{in}, W_{Eh}^{out}$ are the energy input and output of Hteh, respectively; $\phi_{Eh}$ is the heating coefficient. \par
4) Thermal storage tanks (Tsts): the high-temperature Tst receives heat from the furnace top gas $W_{ftg}^{t}$ and the waste heat boiler $W_{whb}^{t}$, and provides heat for preheating the reducing gas. The low-temperature Tst receives heat from the hot hydrogen produced by AE, the compression heat of compressors $\sum\limits_{i = 1}^2 {W_{comp,i}^t}$, and the low-temperature electric heater $W_{leh}^{t}$; its thermal output is the coaxial expander.
\begin{equation}
    Lt{s^{t + 1}} = Lt{s^t} + (W_{Lts,in}^t{\gamma _{in}} - W_{Lts,out}^t/{\gamma _{out}})\Delta t
\end{equation}
\begin{equation}
    Lts^{1} = Lts^{T+1}
\end{equation}
$W_{Lts,in}^{in}, W_{Lts,in}^{out}$ are the thermal inputs and outputs of low-temperature Tst. The high-temperature Tst follows a full-charge/full-discharge operational strategy, while the high-temperature heat balance of the ZCHMS adopts a heat-led operating mode (i.e., heat determines electricity).
\begin{equation}
    W_{Eh}^{in} = W_{sf,{H_2}}^{in} - W_{ftg}^t-W_{whb}^t
\end{equation}\par
5) Constraints on power exchanging with the grid: the ZCHMS is prohibited from purchasing $P_{buy}^{t}$and selling $P_{sell}^{t}$ electricity at the same time. The corresponding constraints are shown as follows:
\begin{equation}
    0 \le P_{sell}^t \le b_{grid}^tM
\end{equation}
\begin{equation}
    0 \le P_{buy}^t \le (1 - b_{grid}^t)M
\end{equation}
$b_{grid}$ is a binary variable; $M$ is a big positive constant.\par
6) Constraints on power balance: the real-time power balance for the ZCHMS is as follows:
\begin{equation}
    P_s^t + P_w^t + P_{buy}^t + P_{\exp }^t = P_{sell}^t + P_{AE}^t + P_{comp}^t + P_{Eh}^t + P_{Leh}^t
\end{equation}
$P_{s}^{t}, P_{w}^{t}$ are the power output of wind and PV, respectively. 
\section{DR potential evaluation framework of ZCHMS}
The DR potential of the ZCHMS is defined as the feasible adjustment range of electricity consumption while satisfying production orders $M_{DRI,order}$. Accordingly, assessing the DR potential amounts to solving a cost-minimization problem that determines the optimal DRI production schedule subject to order-fulfillment constraints.
\subsection{Baseline production scheme}
To ensure that the metallization rate of DRI consistently meets customer specifications, the conventional operational strategy of the ZCHMS maintains a constant discharge flow rate for the SF and a fixed power for the AE. Accordingly, the baseline scheduling scheme and constraints of the ZCHMS are defined as follows:
\begin{equation}
    \begin{aligned}
\min\; \sum_{t=1}^{T}
  Cost_{ma,op}^{t}
  &+ P_{buy}^{t}\,E_{\text{buy,price}}^{t}
  + P_{buy}^{t}\,\phi_{EC}\,C_{\text{tax,price}} \\
  &- P_{sell}^{t}\,E_{\text{sell,price}}^{t}
  - Ml_{sell}^{t}\,Hy_{price}
\end{aligned}
\end{equation}
\begin{equation}
\begin{aligned}
    Cost_{ma,op}^{t}=
    &C^{S}\,P_{s}^{t}+C^{W}P_{w}^{t}+M_{DRI,dis}^{t}\,(C^{Ore}+C^{Sf})+ \\ 
    &C^{AE}\,P_{AE}^{t}+C^{Eh}P_{Eh}^{t}+ C^{coxp}(P_{comp}^{t}+P_{exp}^{t}) \\
    & +2C^{Hs}M_{H_2,in}^{t}+C^{Leh}P_{Leh}^{t}
\end{aligned}
\end{equation}
\begin{equation}
    \sum\limits_{t = 1}^T {M_{DRI,dis}^t = {M_{DRI,order}}} 
\end{equation}
\begin{equation}
    M_{DRI,dis}^t=M_{DRI,QSS}^{K}={M_{DRI,dis}}
\end{equation}
\begin{equation}
    P_{AE}^t = P_{AE}
\end{equation}
$Cost_{ma,op}^{t}$ is the operation cost of ZCHMS; $E_{buy,price}^{t}$ is purchasing electricity prices;$E_{sell,price}^{t}$ is sell electricity prices; $C_{tax,price},Hy_{price}$ are carbon tax and hydrogen price, respectively; $\phi_{EC}$ is the electricity-carbon factor; $C^{i}$ is the operation cost of $i$\text{th} component.
\subsection{DR potential evaluation under real-time pricing}
To match fluctuations in RES and real-time electricity prices, the dynamic flexibility of the SF and the AE in the ZCHMS is utilized. The objective function of ZCHMS under DR scenario is given in eq. (23). The operating load range of the SF and the AE is extended 24\% to 120\% and 10\% to 120\%,respectively; their related constraints are subject to eq. (3)-(5) and eq. (8). The constraint of production orders is formulated as follows:
\begin{equation}
    \sum\limits_{t = 1}^T {M_{DRI,dis}^t = {M_{DRI,order}}}
\end{equation}
\subsection{Evaluation metrics}
To evaluate the DR potential of the ZCHMS under real-time pricing, we adopt the following DR evaluation metrics:
\begin{equation}
    \Delta {P^t} = |P_{DR}^t - P_{baseline}^t|
\end{equation}
\begin{equation}
\overline{{\Delta P}^{t}}
= \frac{\sum_{t=1}^{T} \Delta {P^t}}
       {T}
\end{equation}
$P_{DR}^{t}$ is the power of the ZCHMS under DR scenario; $P_{baseline}^{t}$ denotes the power of the ZCHMS under baseline scheme; $\Delta{P_{t}}$ is the DR potential of the ZCHMS at time $t$; $\overline{{\Delta P}^{t}}$ is the average DR potential of the ZCHMS across the scheduling period. 
\section{Case Study}
The main technical parameters and operation costs of the ZCHMS components are presented in {\color{blue}Table~\ref{1}}. The annual DRI production capacity of SF is set at 1 million tons. The operational conditions are shown in {\color{blue}Fig. 3}.
\begin{figure}[htbp]
\begin{center}
    \setlength{\abovecaptionskip}{-0.1cm}  
    \setlength{\belowcaptionskip}{-0.1cm}   
\includegraphics[width=1\columnwidth]{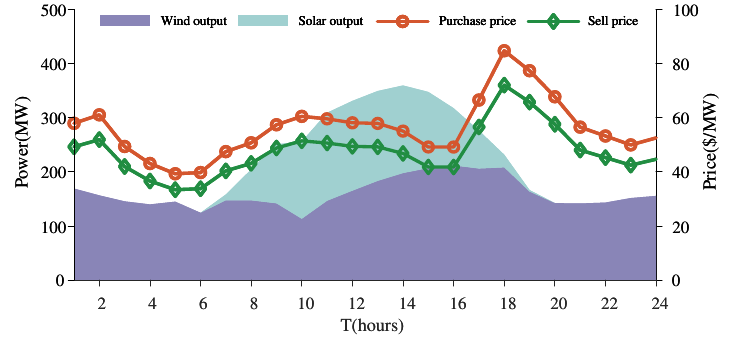}
    \caption{The operational conditions of ZCHMS}
    \label{core diagram}
    \vspace{-0.5cm}
\end{center}
\end{figure}
\begin{table}
\renewcommand{\arraystretch}{1.2}
\begin{center}
\fontfamily{ptm}\selectfont
\footnotesize
\caption{\centering The operational parameters and costs of ZCHMS}
\label{1}
\begin{tabular}{llll}
\toprule
Parameters &Value &Parameters &Value\\ 
\midrule 
$C^{SF}$ & 33.56\$/t & $C^{AE}$ & 13.65\$/MW \\
$C^{coxp}$ & 8\$/MW & $C^{Leh}$ & 4.86\$/MW \\
$C^{Eh}$ & 34.83\$/t & $C^{Ore}$ & 180\$/1.4t \\
$C^{W}$ & 25\$/MW & $C^{S}$ & 30\$/MW \\
$C^{Hs}$ & 6\$/t & $C_{tax,price}$ & 40\$/t \\
$\phi_{EC}$ & 0.57t/MWh & $Hy_{price}$ & 6500\$/t\\
\bottomrule
\vspace{-1cm}
\end{tabular}
\end{center}
\end{table}
\subsection{Operating results of baseline}
To establish a benchmark for comparison, we analyze the baseline load-scheduling strategy of the ZCHMS, characterized by fixed the power of AE and fixed the discharge flow rate of SF; the corresponding results are shown in {\color{blue}Fig. 4}. When PV and wind power outputs peak simultaneously (hours 13-15), the ZCHMS sells electricity to the grid and the state of charge (SoC) of the hydrogen storage tank remains below 0.5. This behavior is driven by the fixed AE power and the hydrogen storage tank's buffering effect, which partially decouples hydrogen production and supply. Under the baseline scheme, the average production cost of DRI is 373 \$/t.
\begin{figure}[htbp] 
    \setlength{\abovecaptionskip}{-0.1cm}  
    \setlength{\belowcaptionskip}{-0.1cm}   
    \includegraphics[width=1\columnwidth]{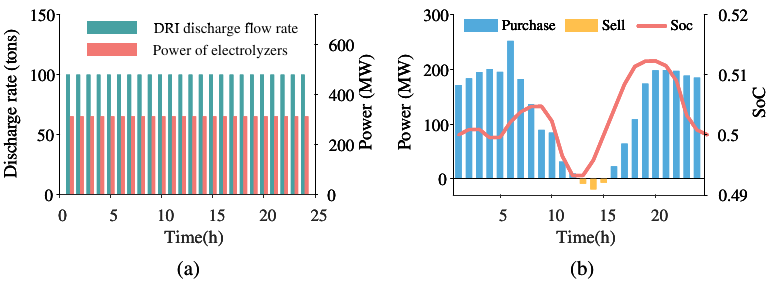}
    \caption{Operation results of baseline. (a) Discharge flow rate of DRI and power of AE; (b) Purchasing/selling electricity and Soc.}
    \label{core diagram}
\end{figure}
\subsection{Operating results of ZCHMS with AE flexibility}
To assess the DR potential of the ZCHMS under flexible AE operation, we analyze its DR strategy and operational costs; the corresponding results are presented in {\color{blue}Fig. 5}. The AE power reaches peak of 720 MW at 03:00 and 24:00, when the ZCHMS purchases 600 MW and 610 MW electricity from the grid, respectively. At 06:00, when both electricity prices and RES outputs are at their lowest, purchasing electricity from the grid peaks at 668MW. During the high-price period of 10:00 to 22:00, RES output fully matches the AE power demand, purchasing electricity from the grid drops to 0, and Soc steadily falls. This behavior illustrates the flexibility of the ZCHMS driven by the AE and the hydrogen storage tank. With this flexible arrangement, the average production cost of DRI reaches 356.7 \$/t, a 4.37\% reduction compared to the baseline. Additionally, the average DR potential of ZCHMS with AE flexibility is 152MW/h.
\begin{figure}[htbp] 
    \setlength{\abovecaptionskip}{-0.1cm}  
    \setlength{\belowcaptionskip}{-0.1cm}   
\includegraphics[width=1\columnwidth]{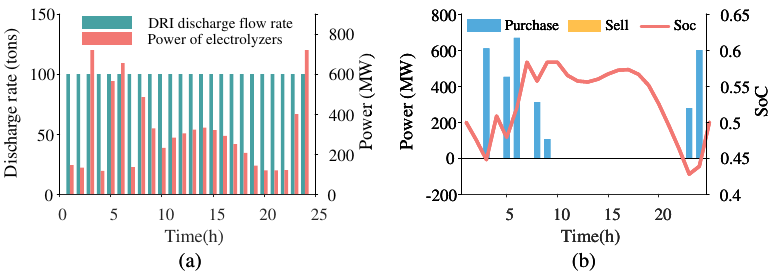}
    \caption{Operation results of ZCHMS with AE flexibility. (a) Discharge flow rate of DRI and power of AE; (b) Purchasing/selling electricity and Soc.}
    \label{core diagram}
\end{figure}
\subsection{Operating results of ZCHMS with AE and SF flexibility}
To evaluate the DR potential of ZCHMS with joint flexibility from both the SF and the AE, we explore  operational strategies of the ZCHMS; the corresponding results are illustrated in {\color{blue}Fig. 6}. From 01:00 to 10:00, the SF runs at its maximum discharge flow rate of 150 t/h. The AE is preferentially loaded between 03:00 and 08:00, matching the low real-time price period and exhibiting price-responsive behavior. At 18:00-19:00, when the selling electricity prices peak, the ZCHMS sells 164 MW and 99 MW to the grid,respectively; meanwhile, the discharge flow rate of SF and power of AE are reduced to their minimum levels of 30 t/h and 60 MW/h. The Soc varies slightly across the scheduling period, which is because SF locks a large circle-hydrogen flow (25.29 t \ce{H2}-150 t/h) to provide the necessary reaction heat. Under this fully flexible configuration, the average production cost of DRI is 348.4 \$/t, and the average DR potential is 179 MW/h.The operating costs and average DR potential of the three schemes are summarized in {\color{blue}Table~\ref{2}}. These results indicate that as the average DR potential rises from 0 to 179 MW/h across the three schemes, the operating cost falls from 373 \$/t to 348.3 \$/t, highlighting the cost reduction value of harnessing flexibility in the ZCHMS.
\begin{figure}[htbp] 
    \setlength{\abovecaptionskip}{-0.1cm}  
    \setlength{\belowcaptionskip}{-0.1cm}    \includegraphics[width=1\columnwidth]{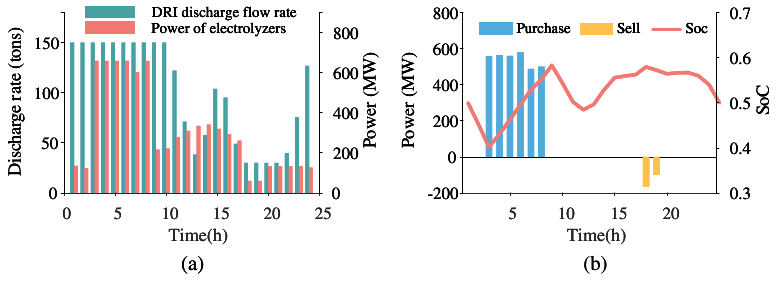}
    \caption{Operation results of ZCHMS with AE and SF flexibility. (a) Discharge flow rate of DRI and power of AE; (b) Purchasing/selling electricity and Soc.}
    \label{core diagram}
    \vspace{-0.5cm}  
\end{figure}
\begin{table}[ht]
\renewcommand{\arraystretch}{1.2}
\begin{center}
\fontfamily{ptm}\selectfont
\footnotesize
\caption{\centering Operating costs and DR potentials across different flexibility scenarios}
\label{2}
\begin{tabular}{lccc}
\toprule
Cases & \textbf{Baseline} & \textbf{AE flex} & \textbf{AE and SF flex}\\
\midrule 
Cost (\$/t)  & 373 & 356.7 & 348.3 \\
Ave DR potential (MW/h) & 0 & 152 & 179 \\
\bottomrule
\end{tabular}
\end{center}
\end{table}
\section{Conclusion}\label{Conclusion}
To narrow the power system flexibility gap, this paper proposes a framework to quantify the DR potential of the ZCHMS. We explicitly model the SF flexibility and validate it on simulation data, achieving a RMSE of 4.48\% of the rated load. The key conclusions are as follows: compared with the baseline scheme, DR-based ZCHMS reduces operating costs by 6.6\%, demonstrating the importance of flexibility to match intermittent renewable energy and real-time electricity prices; and harnessing the flexibility of the ZCHMS strengthens power-steelmaking synergies.
\bibliographystyle{IEEEtran}
\bibliography{Flexibility}

@article{yu2023demand,
  title={Demand response potential evaluation of aggregated high-speed trains toward power system operation},
  author={Yu, Haiyue and Ye, Chengjin and Ding, Yi and Qiu, Lin and Fang, Youtong and Song, Yonghua},
  journal={IEEE Transactions on Smart Grid},
  volume={14},
  number={5},
  pages={3614--3626},
  year={2023},
  publisher={IEEE}
}

@article{li2023coordinated,
  title={Coordinated low-carbon dispatching on source-demand side for integrated electricity-gas system based on integrated demand response exchange},
  author={Li, Chunyan and Yan, Zhichao and Yao, Yiming and Deng, Yulong and Shao, Changzheng and Zhang, Qian},
  journal={IEEE Transactions on Power Systems},
  volume={39},
  number={1},
  pages={1287--1303},
  year={2023},
  publisher={IEEE}
}

@article{golmohamadi2022demand,
  title={Demand-side management in industrial sector: A review of heavy industries},
  author={Golmohamadi, Hessam},
  journal={Renewable and Sustainable Energy Reviews},
  volume={156},
  pages={111963},
  year={2022},
  publisher={Elsevier}
}

@article{lei2023,
  title={Global iron and steel plant CO2 emissions and carbon-neutrality pathways},
  author={Lei, Tianyang and Wang, Daoping and Yu, Xiang and Ma, Shijun and Zhao, Weichen and Cui, Can and Meng, Jing and Tao, Shu and Guan, Dabo},
  journal={Nature},
  volume={622},
  number={7983},
  pages={514--520},
  year={2023},
  publisher={Nature Publishing Group UK London}
}

@article{li2025redesigning,
  title={Redesigning electrification of China’s ammonia and methanol industry to balance decarbonization with power system security},
  author={Li, Jiarong and Lin, Jin and Wang, Jianxiao and Lu, Xi and Nielsen, Chris P and McElroy, Michael B and Song, Yonghua and Song, Jie and Lyu, Xuefeng and Yu, Mingkai and others},
  journal={Nature Energy},
  pages={1--12},
  year={2025},
  publisher={Nature Publishing Group UK London}
}

@article{ji2025energy,
  title={Energy-carbon comprehensive efficiency evaluation of a hydrogen metallurgy system with low-temperature waste heat recovery},
  author={Ji, Qiang and Cheng, Lin and Zhou, Yue and Liang, Zeng and Shi, Fashun and Zhang, Jianliang and Li, Kejiang},
  journal={Applied Energy},
  volume={401},
  pages={126646},
  year={2025},
  publisher={Elsevier}
}

@article{zhang16,
  title={Analysis of process parameters on energy utilization and environmental impact of hydrogen metallurgy},
  author={Zhang, Yujie and Yue, Qiang and Chai, Xicui and Wang, Qi and Lu, Yuqi and Ji, Wei},
  journal={Journal of Cleaner Production},
  volume={361},
  pages={132289},
  year={2022},
  publisher={Elsevier}}

@article{wu2025,
  title={Low-carbon economic dispatch of iron and steel industry empowered by wind-hydrogen energy: Modeling and stochastic programming},
  author={Wu, Haotian and Ke, Deping and Xu, Jian and Song, Lin and Liao, Siyang and Zhang, Pengcheng},
  journal={Applied Energy},
  volume={387},
  pages={125599},
  year={2025},
  publisher={Elsevier}
}

@article{sheng2024,
  title={Rational capacity investment for renewable hydrogen-based steelmaking systems: A multi-stage expansion planning strategy},
  author={Sheng, Kangling and Wang, Xiaojun and Si, Fangyuan and Zhou, Yue and Liu, Zhao and Hua, Haochen and Wang, Xihao and Duan, Yuge},
  journal={Applied Energy},
  volume={372},
  pages={123746},
  year={2024},
  publisher={Elsevier}
}

@article{alarfaj2018material,
  title={Material flow based power demand modeling of an oil refinery process for optimal energy management},
  author={Alarfaj, Omar and Bhattacharya, Kankar},
  journal={IEEE Transactions on Power Systems},
  volume={34},
  number={3},
  pages={2312--2321},
  year={2018},
  publisher={IEEE}
}

@article{otashu2019demand,
  title={Demand response-oriented dynamic modeling and operational optimization of membrane-based chlor-alkali plants},
  author={Otashu, Joannah I and Baldea, Michael},
  journal={Computers \& Chemical Engineering},
  volume={121},
  pages={396--408},
  year={2019},
  publisher={Elsevier}
}

@article{wang2023quantifying,
  title={Quantifying flexibility provisions of the ladle furnace refining process as cuttable loads in the iron and steel industry},
  author={Wang, Jiayang and Wang, Qiang and Sun, Wenqiang},
  journal={Applied Energy},
  volume={342},
  pages={121178},
  year={2023},
  publisher={Elsevier}
}

@article{immonen2024dynamic,
  title={Dynamic modeling of a direct reduced iron shaft furnace to enable pathways towards decarbonized steel production},
  author={Immonen, Jake and Powell, Kody M},
  journal={Chemical Engineering Science},
  volume={300},
  pages={120637},
  year={2024},
  publisher={Elsevier}
}

\end{document}